\documentclass[amsmath,useAMS,usenatbib]{mn2e}
\usepackage{graphicx}
\usepackage{dcolumn}
\usepackage{bm}
\usepackage{color}

\usepackage{verbatim}
\def\lesssim{\stackrel{<}{\sim}}

\begin{document}




 \title{Pseudospectral methods for atoms in strong magnetic fields}
\date{Accepted 2010 April 27.  Received 2010 April 23; in original form 2010 April 2}
\author[J. S. Heyl and A. Thirumalai]{Jeremy
  S. Heyl$^{\dagger}$ and Anand Thirumalai \\
Department of Physics and Astronomy, University of British
Columbia, Vancouver, British Columbia, Canada, V6T 1Z1 \\
$^{\dagger}$Email: heyl@phas.ubc.ca; Canada Research Chair
}

\pagerange{\pageref{firstpage}--\pageref{lastpage}} \pubyear{2010}

\maketitle
\label{firstpage}
\begin{abstract}
  We present a new pseudospectral algorithm for the calculation of the
  structure of atoms in strong magnetic fields.  We have verified this
  technique for one, two and three-electron atoms in zero magnetic
  fields against laboratory results and find typically better than
  one-percent accuracy.  We further verify this technique against the
  state-of-the-art calculations of hydrogen, helium and lithium in strong
  magnetic fields (up to about $2\times 10^{6}$~T) and find a similar
  level of agreement.  The key enabling advantages of the algorithm
  are its simplicity (about 130 lines of commented code) and its speed
  (about $10^2-10^5$ times faster than finite-element methods to
  achieve similar accuracy).  
\end{abstract}
\begin{keywords}
physical data and processes: atomic data --- physical data and
processes: magnetic fields --- stars: neutron --- stars: magnetic
fields --- stars: white dwarfs
\end{keywords}

\section{\label{sec:intro}Introduction}

A hydrogen atom in a magnetic field is arguably the simplest
Hamiltonian without an analytic solution -- it is a combination of a
harmonic oscillator with a Coulomb potential.  The problem of atoms in
magnetic fields is also a vexing challenge in the study of neutron
stars and white dwarfs.  A sufficiently intense magnetic field cannot
be treated perturbatively.  If the magnetic field completely dominates
over the nucleus one can understand the atomic structure as 
a one-dimensional atom \citep{Loud59,Hain69}.   A non-trivial
relaxation of the one-dimensional assumption is the adiabatic
approximation that assumes that the wavefunction of the electron
perpendicular to the magnetic field is simply the ground Landau level
of the electron without the nucleus \citep[e.g.][]{Heyl98atom}.

The most challenging regime is of course where the nucleus and the
magnetic field compete for the electron's attention.  This lies
around $10^{4-6}$~T for hydrogen and is known as the strong-field
regime.  For such field strengths one cannot assume that the electron
lies in the lowest Landau level.  For a detailed bibliography of the vast
literature of atomic calculations in strong magnetic fields, we
encourage the reader to consult our recent work \citep{Thir08atom}.

In the strong-field regime, one can assume that the wavefunction is a
sum of Landau levels or spherical harmonics \citep[e.g.][]{Rude94} or
solve the two-dimensional partial differential equation without this
assumption \citep[e.g.][]{Thir08atom}.  This work takes the second
approach and solves the differential equations for the electronic
wavefunctions and the interelectron potentials numerically.  The main
goal of this paper is not to present more accurate calculations of
atoms in strong magnetic fields (although the calculations rival the
current state of the art), but rather the purpose is to provide a new,
much faster algorithm to perform these calculations and verify its
results against previous work.

This new algorithm is straightforward to implement and compact; the
thirty-line program for single-electron atoms and the 130-line program
for multiple-electron atoms are presented in the appendices.
With this simplicity comes a dramatic speed-up of the calculation of a
factor of $10^2-10^5$ relative to finite-element
techniques \citep{Thir08atom} while achieving similar accuracy.  This
pseudospectral algorithm interpolates the approximate solution to the
differential equations as a polynomial over an irregular mesh.
Because the approximate solution is a polynomial, it is also analytic
over the mesh.  Even if the actual solution oscillates on a scale of a
couple of mesh points, the analytic interpolant still provides an
accurate approximation to the solution.  On the other hand, if the
actual solution develops discontinuities (i.e. it is not analytic),
the pseudospectral method is less useful, \citep[however see][]{Boyd01}.  In the
case of magnetized atoms, the solution to the Schrodinger equation is
analytic everywhere except at the origin where we can set a boundary
condition, so the pseudospectral method yields great returns.

In the next sections we will present the partial differential
equations for the electronic wavefunctions (\S\ref{sec:hamiltonian}),
the interelectron potential (\S\ref{sec:inter-electr-potent}), an
introduction to pseudospectral methods (\S\ref{sec:spect}), the application
of boundary conditions (\S\ref{sec:discretization}), our results
(\S\ref{sec:results}) and source code
(Appendices~\ref{sec:magnetized-hydrogen}-\ref{sec:magnetized-atoms}).

\section{\label{sec:calc}Calculations}

\subsection{\label{sec:hamiltonian}The Hamiltonian}

Rather than examine the Hamiltonian of an atom in a magnetic field in
detail, we will simply recap the relevant equations here and refer the
reader to \citet{Thir08atom} for further details.  We use the
Hartree-Fock equations to describe the wavefunction of a
multiple-equation system, this yields a coupled set of
Schrodinger-like equations for an approximation to the wavefunction of
each electron,
\begin{eqnarray}
\left [ -\nabla^2 - \frac{2}{r_s} + \beta_z^2 R_\perp^2 + \frac{2}{Z}
  \sum_{j,j\neq i} V_{ij}  - E_j \right ] \psi_i = 0
\label{eq:1}
\end{eqnarray}
where $i,j=1,2,3,...,N$.  Here $N$ is the number of electrons, $Z$ is the
charge of the nucleus located at the origin, $V_{ij}$ is the
inter-electron potential operator and $\beta_Z$ measures the strength
of the magnetic field, $\beta_Z = B e\hbar/(2 Z^2 \alpha^{2}
m_{e}^2c^2)=B/(4.7 Z^2 \times 10^5 {\rm T})$.  The variable $r_s$ is
the distance from the origin and $R_\perp=r_s\sin\theta$, the distance
from the axis defined by the direction of the magnetic field; this
notation is consistent with the software in the appendices.  The
functions $\psi_i$ must be finite at the origin and approach zero as
${\bf r} \rightarrow \infty$.

The total binding energy of the atom is given by
\begin{equation}
E = \sum_i \left [ E_i + 2 \beta_Z \left (m_i + s_i - \frac{1}{2} \right )
 - \frac{1}{Z} \sum_{j,j\neq i} \langle \psi_i |V_{ij}|
 \psi_i\rangle \right ]
\label{eq:2}
\end{equation}
in units of $Z^2 \alpha^2 m_e c^2/2$

Although the Hamiltonian is no longer spherically symmetric, it is
still symmetric about the magnetic axis, so we expect the azimuthal or
magnetic quantum number $m_i$ to be good.  We will label the solutions
to Eq.~(\ref{eq:1}) by $\nu_i, s_i$ and $m_i$ where $s_i$ is the spin of
the electron and $\nu_i$ equals 1 for the most bound state with a
particular value of $s_i$ and $m_i$ and 2 for the next most bound
state and so on. 

The first step is to specify the precise differential equations to
solve.  Specifically, we will take
\begin{equation}
\psi_i = \frac{u_i(r,\mu)}{r_s} e^{i m_i \phi}.
\label{eq:3}
\end{equation}
The parameter $\mu=\cos\theta$, where $\theta$ is the angle from the z-axis. For the radial dependence, the domain is $[0,\infty)$, so we would like
compactify the domain to $[-1,1]$.  Let
\begin{equation}
r = 2 \tanh \frac{r_s}{r_Z} - 1
\label{eq:4}\end{equation}
where $r_s$ ranges from zero to infinity  while $r$ now ranges from $-1$
to 1.  The quantity $r_Z$ is the ``zoom radius'' that determines at what
value of $r_s$ the relationship between $r$ and $r_s$ goes from linear to
exponential,
\begin{equation}
r \approx 2 \frac{r_s}{r_Z} -1, r\ll r_Z; r \approx 1-4\exp(-2 r_s/r_Z),
r\gg r_Z.
\label{eq:5}\end{equation}
Because $2\tanh 1 - 1\approx 0.5$ about one third of the points lie at
$r>r_Z$.  Because the domain is compactified we can easily apply boundary
conditions both at infinity and at the origin as appropriate for
solving the Schrodinger equation: $u_i(-1,\mu)=u_i(+1,\mu)=0$ and the
additional condition for $m_i\neq 0$ that $u_i(r,-1)=u_i(r,+1)=0$.

Using the substitution of Eq.~(\ref{eq:3}) into~(\ref{eq:1}), yields
\begin{eqnarray}
\Biggr [  -\frac{\partial^2}{\partial r_s^2} - \frac{1}{r_s^2} \left ( 1 - \mu^2 \right
  ) \frac{\partial^2}{\partial \mu^2} + \frac{2\mu}{r_s^2}
  \frac{\partial}{\partial \mu} + \frac{m_i^2}{R_\perp^2} -
  \frac{2}{r_s} ~~~~
\nonumber 
 \\
 + \beta_Z^2 R_\perp^2 
 + \frac{2}{Z}   r_s e^{-i m_i \phi} \sum_{j,j\neq i} V_{ij}
  \frac{e^{im_i \phi}}{r_s} - E_j \Biggr ] u_i = 0  \label{eq:6}
\end{eqnarray}
where
\begin{equation}
\frac{\partial^2}{\partial r_s^2} = 4 \left ( 1 - r_p^2  \right )
\left [ \left ( 1 - r_p^2 \right ) \frac{\partial^2}{\partial r^2} -
  r_p \frac{\partial}{\partial r} \right ]\label{eq:7}
\end{equation}
where $r_p=(r+1)/2$.

\subsection{\label{sec:inter-electr-potent}Inter-electron potential}

Using the results of \S\ref{sec:spect} we can convert this partial
differential equation into a matrix equation, specifically an
eigenvalue equation.  However, before doing this in
\S\ref{sec:discretization}, we must discuss the inter-electron
potential $V_{ij}$.

The inter-electron potential consists of two terms, the direct and
exchange interactions where
\begin{equation}
V_{ij} = V_{{\rm direct},ij}  - V_{{\rm exchange},ij} \label{eq:8}
\end{equation}
if the spins of electrons $i$ and $j$ are aligned and $V_{ij} =
V_{{\rm direct},ij}$ otherwise. The direct interaction satisfies
\begin{equation}
V_{{\rm direct},ij} \psi_i = \phi_j({\bf r}) \psi_i
\label{eq:9}\end{equation}
where
\begin{equation}
\nabla^2 \phi_j =  -4\pi \psi_j^* \psi_j
\label{eq:10}\end{equation}
which we can solve, knowing the values of $\psi_j$, by inverting the
Laplacian to yield
\begin{equation}
 \phi_j =  \left ( \nabla^2 \right )^{-1} \left [ -4\pi \psi_j^*
   \psi_j \right ]
\label{eq:11}\end{equation}
where $\left ( \nabla^2 \right )^{-1}$ is the inverse of the Laplacian
with the appropriate boundary conditions supplied.  Therefore, the
direct interaction is a diagonal operator, essentially a function.

On the other hand the exchange interaction is more complicated.  We
have
\begin{equation}
V_{{\rm exchange}, ij} \psi_i = \tilde\phi_{ij}({\bf r}) \psi_j
\label{eq:12}\end{equation}
where
\begin{equation}
\nabla^2 \tilde\phi_{ij} = -4\pi \psi_j^* \psi_i .
\label{eq:13}\end{equation}
Therefore, we have
\begin{equation}
V_{{\rm exchange}, ij} \psi_i = \tilde\phi_{ij}({\bf r}) \psi_j 
= \psi_j \left ( \nabla^2 \right )^{-1} \left [ -4\pi \psi_j^*
   \psi_i \right ].
\label{eq:14}\end{equation}
Thinking of the different components of the right-hand sides of Eqs.~(\ref{eq:12}) and (\ref{eq:15}) as matrices, we can write the potentials due to the direct and exchange interactions in matrix notation as
\begin{equation}
V_{{\rm exchange}, j} = -4\pi~{\rm diag} (\psi_j) \left ( \nabla^2
\right )^{-1}
 {\rm diag} (\psi_j^*) 
\label{eq:15}\end{equation}
where we have dropped the index $i$ because the exchange interaction
operator only depends on the wavefunction of the electron $j$.
Both the direct and exchange potentials must go to zero at ${\bf r}
\rightarrow \infty$ and be finite at the origin. 

The set of coupled equations is of course non-linear and difficult to
solve directly; however, we can approach the solution iteratively, by
first solving the equations~(\ref{eq:1}) ignoring the inter-electron
terms and then on subsequent iterations using the wavefunctions from
the previous iteration to calculate the interaction operator using
Eq.~(\ref{eq:11}) and~(\ref{eq:15}).  At each iteration the equations
are decoupled, so the time to calculate a single iteration scales as
the number of electrons, making the calculation of even large atoms
tractable.  If the equations were coupled through the exchange term,
then each dimension of the matrix to diagonalize would scale with the
number of electrons so the time to complete an iteration would scale
as the cube of the number of electrons.

Let's evaluate for the exchange potential operator using the
substitution of Eq.~(\ref{eq:3}).  We have
\begin{equation}
V_{{\rm exchange}, ij} \left ( u_i \frac{e^{i m_i \phi}}{r_s} \right )  =
\tilde\phi_{ij}({\bf r}) u_j \frac{e^{i m_j \phi}}{r_s}
\label{eq:16}
\end{equation}
and
\begin{equation}
r_s e^{-i m_i \phi} V_{{\rm exchange}, ij} \left ( \frac{e^{i m_i \phi}}{r_s}
u_i \right )  = \tilde\phi_{ij}({\bf r}) u_j e^{-i \Delta m \phi}
\label{eq:17}
\end{equation}
where $\Delta m=m_i-m_j$.  Because none of the other terms in the
Hamiltonian depend of $\phi$, neither side of Eq.~(\ref{eq:17})
can depend on $\phi$ so let us define
\begin{equation}
\tilde\phi_{ij}({\bf r}) = \tilde\phi^{\Delta m}_{ij}(r_s,\mu) e^{i \Delta m \phi}
\label{eq:18}\end{equation}
where
\begin{equation}
\nabla^2 \left ( \tilde\phi^{\Delta m}_{ij}(r_s,\mu) e^{i \Delta m
  \phi} \right ) = -4\pi \frac{u_j^* u_i}{r_s^2} e^{i\Delta m \phi}.
\label{eq:19}
\end{equation}
Simplifying this result yields
\begin{equation}
\left ( \nabla^2 - \frac{\Delta m^2}{R_\perp^2} \right )
\tilde\phi^{\Delta m}_{ij}(r_s,\mu) = -4\pi \frac{u_j^* u_i}{r_s^2}\label{eq:20}
\end{equation}
and
\begin{eqnarray}
r_s && \!\!\!\!\!\!\!\!\!\!\!\! e^{-im_i\phi}  V_{{\rm exchange}, ij} \frac{e^{im_i\phi}}{r_s} 
 \nonumber \\
 && = -4\pi~{\rm diag} (u_j) \left ( \nabla^2 - \frac{\Delta
  m^2}{R_\perp^2} \right )^{-1} 
 {\rm diag} \left ( \frac{u_j}{r_s^2} \right ) 
\label{eq:21} 
\end{eqnarray}
where
\begin{equation}
 \nabla^2   =
\frac{\partial^2}{\partial r_s^2} 
 + \frac{1}{r_s^2} \left ( 1 - \mu^2 \right
   ) \frac{\partial^2}{\partial \mu^2} - \frac{2\mu}{r_s^2}
   \frac{\partial}{\partial \mu} .
\end{equation}

The result for the direct interaction is nearly trivial because it is
a function rather than an operator.  We have
\begin{equation}
r_s e^{-i m_i \phi} V_{{\rm direct}, ij} \left ( \frac{e^{i m_i \phi}}{r_s}
u_i \right )  = -4\pi  \left ( \nabla^2 \right )^{-1} \left [ \frac{u_j^2}{r_s^2}
  \right ]
\label{eq:23}
\end{equation}
We must specify the boundary conditions on the inverse
operators in Eq.~(\ref{eq:21}) and~(\ref{eq:23}). 
Both the direct and exchange potentials must go to zero at ${\bf r}
\rightarrow \infty$ and be finite at the origin. Furthermore, for
$\Delta m \neq 0$, the exchange potential must vanish along the
magnetic axis.

In principle one could solve for all of the electronic wavefunctions
at the same time, but this would render the differential equations
non-linear; therefore, we first neglect the electronic contribution to
the potential to find a first guess at the electronic wavefunctions.
These wavefunctions become the input into Eq.~(\ref{eq:21})
and~(\ref{eq:23}) for the second iteration.  This process is repeated
until the total energy of the state is constant to one part in $10^5$
between successive iterations.

\subsection{\label{sec:spect}Pseudospectral methods}

The basic idea behind interpolating spectral (or pseudospectral)
methods is to approximate functions by their values at a set of points
and a set of smooth functions to interpolate between these points for
the purpose of calculating derivatives and integrals.  An example is
useful to make this concrete.  A function $f(x)$ takes the values
$f_i$ at the points $x_i$.  We could construct an interpolation scheme
by linearly interpolating between the points or using a cubic spline.
This is the basis of finite difference schemes with successively
higher accuracy.  However, if we wanted an analytic function through
the points we could use the polynomial that passes through all of the
points.  This polynomial interpolant is given by
\begin{equation}
f_I(x) = \sum_i f_i p_i(x) = \sum_i f_i \frac{\prod_{j\neq i} (x-x_j)}{\prod_{j\neq i} (x_i-x_j)}
\label{eq:24}
\end{equation}
where we have used the notation of \citet{Tref00}.  The polynomial
$p_i(x)$ has the property that it takes the value of unity at $x_i$
and it vanishes at all of the other points where the function values
are given.  We can take the derivative of the basic polynomial to
yield
\begin{equation}
p_i^\prime(x) = p_i(x) \sum_{j\neq i} (x-x_j)^{-1}
\label{eq:25}
\end{equation}
and evaluate it at one of the points $x_k$ to yield
\begin{equation}
p_i^\prime(x_k) = p_i(x_k) \sum_{j\neq i} (x_k-x_j)^{-1} = D_{ik}
\label{eq:26}
\end{equation}
where
\begin{equation}
D_{ik} = \frac{\prod_{j\neq i,k} (x_k-x_j)}{\prod_{j\neq i} (x_i-x_j)}, D_{ii} = \sum_{j\neq i} (x_i-x_j)^{-1}
\label{eq:27}\end{equation}
where $D_{ik}$ is the differentiation matrix such that 
$f_I^\prime(x_k) = D_{ik} f_i$.   Unlike finite element, finite volume
and finite difference techniques, the differentiation matrix is dense,
so it is more costly to manipulate; however, one can often achieve
similar accuracy with much smaller matrices for problems with smooth
solutions.  This decrease in the size of the matrices more than
offsets the burden of handling dense matrices.

The choice of the points $x_i$ is completely arbitrary but for finite
domains it is more stable to pick the points to be bunched toward the ends
of the domain and more sparse in the middle \citep{Tref00}.  A
convenient set of points over the domain $[-1,1]$ are the Chebyshev
points 
\begin{equation}
x_i = \cos ( i \pi / N ), i=0\ldots N.
\label{eq:28}\end{equation}
This domain is well suited for the angular dependence in spherical
coordinates where we define $\mu=\cos \theta=z/r$ and use the
Chebyshev points to sample $\mu$.  Thus the angular dependence is
evenly sampled in $\theta$.

\subsection{Discrete equations\label{sec:discretization}}

The {\tt Octave} programs given in the appendices for the single and
multiple-electron problems convert the partial differential equations
of \S\ref{sec:hamiltonian} and~\ref{sec:inter-electr-potent} to
matrix equations using the pseudospectral methods outlined in
\S\ref{sec:spect}.  We encourage readers that are familiar with {\tt
  Octave} or {\tt Matlab} to look under the hood and dissect the
routines.

The previous section outlined how to construct a differentiation
matrix using a pseudospectral method in one dimension.  The second
derivative is simply the square of this matrix and an integration
operator is the inverse of the differentiation matrix with some
boundary condition supplied.  The remaining subtleties are how to
apply boundary conditions and how to move to more than one
dimension. The techniques outlined here are presented in
\citet{Tref00} in greater detail, clarity and generality.

Supplying the boundary conditions for the eigenvalue equations is
rather straightforward and most easily illustrated by an example.
Let us find the eigenvalues of the matrix $M$ with the solution $f$
vanishing at the end points.  We have
\begin{equation}
M f = \lambda f.
\label{eq:29}\end{equation}
Let's define the diagonal matrix,
\begin{equation}
B = \left [ \begin{array}{cccccc} 
0 & 0 & 0 & 0 & \cdots & 0 \\
0 & 1 & 0 & 0 & \cdots & 0 \\
0 & 0 & 1 & 0 & \cdots & 0 \\
\vdots & & & \ddots & & 0 \\
0 & 0 & 0 & \cdots & 1 & 0 \\
0 & 0 & 0 & \cdots & 0 & 0 
\end{array} \right ]\label{eq:30}
\end{equation}
with all of the diagonal elements unity except for the first and last that
vanish.  Multiplying the vector $\vec{f}$ by $B$ has the effect of setting the
endpoints to zero, so let's look at the following eigenvalue equation
\begin{equation}
B M B \vec{f} = \lambda \vec{f}.\label{eq:31}
\end{equation}
The matrix $B$ that precedes $M$ ensures that the resulting output
vector satisfies the boundary conditions, and the matrix that follows
$M$ ensures that the input vector satisfies the boundary conditions as
well, so the eigenvectors of the matrix $BMB$ except for two are
guaranteed to satisfy the boundary conditions.  The two remaining
eigenvectors are linear combinations of $\left [ 1\, 0\, \cdots\, 0\,
  0 \right ]$ and $\left [ 0\, 0\, \cdots\, 0\, 0\, 1 \right ]$ and
have zero eigenvalues.  We can either ignore these eigenvectors or
define a new matrix $\tilde M$ that lacks the first and last columns
and the first and last rows of the original matrix $M$.  The
eigenvectors of $\tilde M$ are the same at those of $B M B$ but with
the first and last entries omitted.

In summary, to obtain the set of eigenvectors of $M$ that satisfy the
boundary conditions we construct $\tilde M$ and add the first and last
entries if needed.  For the potential calculation it is often more
useful to work with $B M B + (I - B)$ so that the resulting matrix can
be inverted before setting the boundary conditions.  In the potential
calculation we set the boundary condition at infinity because the
Laplacian matrix vanishes for the row and column corresponding to
infinity, so we drop the row and column before inverting the
Laplacian.  However, we would also like to set a boundary condition at
the origin, so we must include the origin in the calculation.  At the
origin the angular terms in the Laplacian diverge, but this is a
coordinate singularity that we address by including only a single
entry for the origin and drop the angular derivatives there.  In this
way, we address the divergent terms at the origin and the vanishing
terms at infinity.

The second issue is how to build the various operators in more than
one dimension.  Let's say that we have obtained a vector of points
$\vec x$ of dimension $N$ and a pseudospectral derivative $D_x$ and
similarly a vector of points $\vec y$ of dimension $M$ and its
derivative $D_y$.  If we construct
\begin{equation}
\vec x' = \vec x \otimes \vec 1_M, \vec y' = \vec 1_N \otimes \vec y 
\label{eq:32}\end{equation}
where $\vec 1_N$ is a vector of $N$ ones and $\otimes$ denotes the
Kronecker product, we obtain a two-dimensional mesh $(\vec x',\vec
y')$ of points in two vectors of dimension $N
\times M$.  Finally we would like to construct the Laplacian over this
mesh (this example is Cartesian).  We have
\begin{equation}
\nabla^2 = (D_x)^2 \otimes I_M + I_N \otimes (D_y)^2.
\label{eq:33}\end{equation}
where $I_N$ is the $N\times N$-identity matrix.  The resulting matrix
$\nabla^2$ has dimensions $(N\times M) \times (N\times M)$ and
calculates the Laplacian over the mesh specified by $(\vec x',\vec
y')$.

These techniques allow us to discretize the equations outlined in
\S\ref{sec:hamiltonian} and~\ref{sec:inter-electr-potent} to obtain
the software in the appendices.

\section{Results}
\label{sec:results}

\subsection{Hydrogen}

As a first test of the algorithm, we examined hydrogen in a strong
magnetic field and in zero field.  The {\tt Octave} program is
presented in Appendix~\ref{sec:magnetized-hydrogen}.  This is one case
where the program is shorter than the tables it calculates and
replaces.  In zero-field with 30 radial points ($N=31$, excluding
excluding $-1$ and $1$ in the compactified domain, corresponding to
the origin and infinity, respectively in the real domain) and with 12
angular points ($M=11$), it calculates the eigenvalues up to principal
quantum number, $n=12$ to at least three digits of accuracy (a total
of 650 states).  Up to $\beta\sim 10$ it determines the low-lying
states of hydrogen (again with $N=31$ and $M=11$) to at least three
digits of accuracy \citep{Thir08atom,Rude94}; for $\beta \lesssim 1$,
the accuracy is typically two orders of magnitude better.  The
distinct advantage of the pseudospectral algorithm is the speed to
achieve accurate results.  The fine mesh calculations of
\citet{Thir08atom} took up to two days or $1.7 \times 10^5$ seconds
--- the pseudospectral algorithm takes about 1.4~s to get the same
accuracy on the same processors.  The algorithm uses spherical
coordinates so it is not well suited to fields much stronger than
$\beta=10$.  At $\beta=100$, a finer mesh ($M=31$ and $N=51$) and 150
seconds of computation time are required to obtain three digits of
agreement with \citet{Rude94}; however, in this regime, cylindrical
coordinates may be more appropriate and one can use the adiabatic
approximation to obtain accurate results \citep{Heyl98atom}.

\subsection{Helium and Lithium}
The problem of multi-electron atoms
(Appendix~\ref{sec:magnetized-atoms}) is of course much more
complicated than hydrogen.  Again we begin with the zero-field case,
but we have many states to examine that test the direct and exchange
interactions separately.  In this case some validation for a vanishing
magnetic field is useful.  Table.~\ref{tab:heli} compares the
zero-field results for some low-lying states of helium and lithium with
the experimentally determined values from \citet{Ralc08}.  Again the algorithm
achieves typically better than one-percent accuracy rapidly.
\begin{table}
  \caption{Comparison of the observed zero-field states of helium and
    lithium from
    \citet{Ralc08} to the values calculated here. 
    Energies are in units of Rydberg energies in the Coulomb potential of
    nuclear charge $Ze$, where $Z=2$ for helium and $Z=3$ for lithium. We use $M=7,N=31$
    and $r_Z=58$.  The value of $r_Z$ is chosen to provide close agreement
    for the ground state of helium and held fixed.  Individual calculations take
    about ten seconds for helium and twenty for lithium.  The primary 
    source for the helium data is \citet{1998CaJPh..76..679D} and
    \citet{Moor71} for the lithium data.}
\label{tab:heli}
\begin{tabular}{cccrrr}
\hline
Conf. & Term & $J$ & \multicolumn{1}{c}{$E_{\rm exp}$} &
\multicolumn{1}{c}{$E_{\rm spe}$} & \multicolumn{1}{c}{$\Delta$(\%)} \\
\hline 
\multicolumn{6}{c}{Helium I} \\
\hline 
$1s^2$ & $^1\!S$      & 0 & 1.451692958 & 1.43189& -1.364 \\
$1s2s$ & $^3\!S$      & 1 & 1.087514256 & 1.08996& 0.225 \\
$1s2s$ & $^1\!S$      & 0 & 1.072885081 & 1.07929& 0.598 \\
$1s2p$ & $^3\!P^\circ$ & 2 &1.066484964 & 1.06856 & 0.195 \\
       &              & 1 & 1.066484790 & 1.06857& 0.196 \\
       &              & 0 & 1.066482539 & 1.06856& 0.195 \\
$1s2p$ & $^1\!P^\circ$ & 1 & 1.061818980 & 1.06593 & 0.387 \\
       &              & 1 & 1.061818980 & 1.06594& 0.388 \\
$1s3s$ & $^3\!S$      & 1 & 1.034248834 & 1.03719& 0.284 \\
$1s3s$ & $^1\!S$      & 0 & 1.030539891 & 1.03456& 0.391 \\
$1s3p$ & $^3\!P^\circ$ & 2 & 1.028945784 & 1.03174 & 0.272 \\
       &              & 1 & 1.028945735 & 1.03174& 0.271 \\
       &              & 0 & 1.028945117 & 1.03174& 0.272 \\
$1s3d$ & $^3\!D$      & 3 & 1.027722445 & 1.03073& 0.293 \\
       &              & 2 & 1.027722439 & 1.03073& 0.293 \\
       &              & 1 & 1.027722338 & 1.03073& 0.293 \\
$1s3d$ & $^1\!D$      & 2 & 1.027714652 & 1.03073& 0.294 \\
$1s3p$ & $^1\!P^\circ$ & 1 & 1.027476815 & 1.03092 & 0.335  \\
\hline 
\multicolumn{6}{c}{Lithium I} \\
\hline 
$1s^22s$ & $^2\!S$      & $1/2$ & 1.661773620 & 1.65595& -0.350 \\
$1s^22p$ & $^2\!P^\circ$ & $1/2$ & 1.646683387 & 1.64097 & -0.347 \\
        &               & $3/2$ & 1.646683042 & 1.64097& -0.347 \\
$1s^2 3s$ & $^2\!S$     & $1/2$ & 1.634226909 & 1.62883& -0.330 \\
\\
\end{tabular}
\end{table}

Of course, the technique is not intended to supplant the popular,
accurate and fast MCHF algorithms such as those of
\citet{CFF1997}.  Such algorithms achieve their speed and accuracy
by using the spherical symmetry of the atom: for example, by using
spherical harmonics and a special treatment of full shells.  The power
of this pseudospectral algorithm becomes crucial when spherical symmetry
is destroyed by a strong magnetic field.
\begin{table}
\centering
\caption{Table listing the binding energies of the most tightly bound
  states of helium in moderate to large magnetic fields; $M_z=-1$ or
  $M_z=-2$ and $S_{z}=-1, \pi_{z}=+1$. Energies are in units of Rydberg energies in
  the Coulomb potential of nuclear charge $Ze$, where $Z=2$ for
  helium. The results from the current work can be compared readily
  with previous work \citep{Thir08atom,Rude94}.  The first row for
  $\beta_Z=1$ is for $M=11$ like the other rows; the second row gives
  the results from $M=41$.  TH08 indicates results from
  \citet{Thir08atom}, and R94 indicates results from \citet{Rude94}.
}
\begin{tabular}{c|lcc|ccc}
\hline
\hline
& \multicolumn{3}{c|}{$M_z=-1$} & \multicolumn{3}{c}{$M_z=-2$} \\
$\beta_{Z}$ & Here & TH08 &  R94 & Here &
TH08 & R94 \\
\hline
0.01 & 1.1221 & 1.1183 & 1.1182& 1.0871 &  1.0852 & 1.0828 \\
0.02 & 1.1656 & 1.1612 & 1.1609& 1.1265 &  1.1234 & 1.1207 \\
0.05 & 1.2734 & 1.2691 & 1.2658& 1.2215 &  1.2175 & 1.2097 \\
0.07 & 1.3359 & 1.3319 & 1.3258& 1.2764 &  1.2732 & 1.2596 \\
0.10 & 1.4211 & 1.4189 & 1.4069& 1.3515 &  1.3510 & 1.3266 \\
0.20 & 1.6586 & 1.6585 & 1.6270& 1.5653 &  1.5598 & 1.5073 \\
0.50 & 2.1603 & 2.1550 & 2.0508& 2.0415 &  2.0009 & 1.8508 \\
0.70 & 2.4584 & 2.4029 & 2.2329& 2.2902 &  2.2246 & 1.9935 \\
1.00 & 3.2165 & 2.7026 & 2.4675 & 3.0102 &  2.4981 & 2.2572 \\
1.00 & 2.7115 &        &        & 2.6668 &          &        \\
\hline
\hline
\end{tabular}
\label{tab:hem-1}
\end{table}

Table.~\ref{tab:hem-1} traces the binding energy of two low-lying states
of helium using the software in Appendix~\ref{sec:magnetized-atoms}.
The agreement with our earlier work \citep{Thir08atom} is striking at
weaker fields but as the magnetic field starts to dominate the atom,
the agreement is poorer.  Better agreement is achieved by increasing the
angular resolution of the calculation to account for the changing
shape of the atom.

For lithium we compare with the results of
\citet{Schmelcher_lithium2004} who use a different parameter for the
strength of the field, $\gamma=2 Z^2\beta_Z=18\beta_Z$ where the
second equality holds for lithium.  We have chosen to use eight
angular grid points and thirty-two in the radial direction.

\begin{table}
\centering
\caption{
  Table listing the binding energies of three tightly bound
  states of lithium in moderate to large magnetic fields. At the top
  of each pair of columns is the symmetry subspace
  $^{2S+1}(M_z)^{\rm z-parity}$.  We take the
  dominant configuration for the ground state within each subspace.
  The column ``Other'' gives the values from \citet{Ralc08} for
  zero field and from \citet{Schmelcher_lithium2004} elsewhere.
}
\begin{tabular}{c|cc|cc|cc}
\hline
\hline
$\gamma$
& \multicolumn{2}{c|}{$^20^+$} &
\multicolumn{2}{c|}{$^4(-1)^+$} &
\multicolumn{2}{c}{$^4(-3)^+$} \\
$18\beta_{Z}$ & Here & Other &  Here & Other & Here & Other \\
\hline  
0.00 &   1.6560 & 1.6618 & 1.1968 & 1.1925 &  1.1358 & 1.1299 \\
0.01 &   1.6590 & 1.6629 & 1.2017 & 1.1969 &  1.1427 & 1.1487 \\
0.05 &   1.6681 & 1.6673 & 1.2192 &  ---   &  1.1655 & 1.1663 \\
0.10 &   1.6775 & 1.6705 & 1.2390 & 1.2334 &  1.1901 & 1.1869 \\
0.20 &   1.6928 & 1.6741 & 1.2736 & 1.2674 &  1.2330 & 1.2278 \\
0.50 &   1.7255 & 1.6729 & 1.3533 & 1.3463 &  1.3362 & 1.3294 \\
1.00 &   1.8256 & 1.6575 & 1.4821 & 1.4432 &  1.4713 & 1.4627 \\
\hline            
\hline
\end{tabular}
\label{tab:lim-1}
\end{table}

Here the agreement with the previous work is encouraging at least up
to moderate field strengths ($\gamma < 1$).  For stronger fields the
pseudospectral method typically yields excessively bound systems,
except for the $1^4(-3)^+$ state that happens to be the most bound
state in strong magnetic fields.  These spurious eigenvalues are not
uncommon for pseudospectral methods \citep[e.g.][]{Boyd01}.  The
software as presented in the appendices actually includes a filter for
a blatantly spurious eigenfunction that attains a large value along
the magnetic axis for large distances from the nucleus.  It is our
experience with the hydrogen atom that these spurious eigenvalues
become more prevalent for stronger magnetic fields and are more bound
than the eigenvalues for well behaved eigenfunctions.  However, the
algorithm still finds the correct eigenvalue for the ground state, so
with a more sophisticated filtering technique spurious eigenvalues can
be eliminated for calculations in higher field strengths than
considered here.

\section{\label{sec:discussion}Discussion}

One of the primary goals of the current work was to provide a
computational method that would be rather economical with regard to
computation time, without having to compromise on accuracy.  \ As can
be seen in Table~\ref{tab:heli} the desired level of accuracy can be
achieved with computation times on the order of seconds. The
calculated values therein can conceivably have better agreement with
experimentally determined values if effects of spin-orbit coupling,
relativistic corrections and the effects of mixing of configurations
were to be included. Similarly, upon examining Table~\ref{tab:hem-1},
it is immediately apparent that the estimates of the binding energies
of the state of helium corresponding to $M_z=-1; S_z=-1$ are more
bound than the previous estimates. The average improvement with
respect to the values reported in our previous study
\citep{Thir08atom}, is about 0.5\% while, the average improvement over
the values reported in \citet{Rude94} is about 3.4\% over the
entire range of magnetic field strengths reported therein. Similarly,
upon comparing the results from the present calculation for the state
of helium corresponding to $M_z=-2; S_z=-1$ we see that they are more
bound on average by about 1.5\% and 5.8\% respectively, with respect
to the calculations of \citet{Thir08atom} and
\citet{Rude94}. Finally, when comparing with the results of other
authors in Table~\ref{tab:lim-1} we see that the current
single-configuration calculation using spectral methods provides
binding energies that are an improvement on previous estimates for three
tightly bound states of lithium. It can be seen that the average range
of improvements for the estimates of the binding energies of the three
states, relative to the calculations of other work, is about 1\%.

In carrying out the calculations in the current study, as was noted
earlier, the domain was compactified according to
Eq.~(\ref{eq:4}). This has the distinct advantage that one can set a
boundary condition at true infinity rather than at a large value of
the radial distance $r$. Typically, such a procedure renders the
partial differential equations highly non-linear, making their
solution with finite-element techniques more involved and
computationally demanding, thus requiring a greater amount of
computation time. In the current study, the compactification
effectively changes the operators in Eqs.~(\ref{eq:6}), (\ref{eq:21})
and (\ref{eq:23}).   

Since the algorithm presented here is essentially a simple prototype,
several avenues for further development are apparent.  They address
the stability, accuracy and robustness of the calculation.  First from
a physical point of view it does not make sense to use the same zoom
radius (the value of $r_Z$) for all of the electrons in an atom.  The
zoom radius is essentially the scale within which the radial mesh is
fine; therefore, it should reflect the expected or calculated extent
of the electronic wavefunctions.  It would be quite natural to
adaptively change the value of $r_Z$ to be some constant factor times
$\langle r \rangle$ or $\sqrt{\langle r^2 \rangle}$ as the
Hartree-Fock iteration proceeds.  Because the interelectron potential
is smoother than the underlying wavefunctions it would be more
effective to calculate the potentials for each electron using its own
value of $r_Z$ and then interpolate these results onto the meshes of
the other electrons.  In this manner the value of $r_Z$ would adapt
for the particular orbital that the electron occupies, achieving
higher accuracy without increasing the number of radial mesh points.
For the direct interaction this is straightforward because it is
essentially a function; however, the exchange interaction is an
operator or a matrix, as in Eq.~(\ref{eq:15}), so the interpolation of
the final result is a bit problematic. Rather it is expedient to
interpolate the wavefunctions of the other electrons onto each other's
mesh and calculate the exchange interaction one pair at a time.
Although this interpolation would necessarily make the code more
cumbersome, it need not increase the processing time significantly.
The inverse Laplacians for different values of $r_Z$ are related to
each other through simple scalings so the costly matrix inversions
would still only happen once at the beginning, and the results would
scale for the appropriate values of $r_Z$.

A similar adaptation can be performed in the angular direction to
focus grid points along the magnetic axis for strong magnetic fields.
In this way we can achieve high accuracy for all of the electrons
without including more grid points.  A simple way to achieve this is
the ``Arctan/Tan'' mapping \citep{Boyd01} where the original angular
variable $\theta$ is mapped onto $\theta'$ using
\begin{equation}
\theta' = \arctan \left ( L \tan \theta \right)
\end{equation}
The parameter $L$ can change to reflect the actual physical extent of
the wavefunction in the angular direction for each electron in the
atom as the iterative process proceeds.  This adaptation would add little
additional overhead above the radial adaptive scheme outline in the
last paragraph.

A second issue that we encountered is the appearance of spurious
eigenvalues among the physical ones, especially for intense magnetic
fields.  The current software eliminates these eigenvalues by looking
at the structure of the eigenfunctions.  Essentially, many of the
spurious eigenfunctions have significant support along the magnetic
axis at large radii.  Unfortunately, the current set of criteria are
not sufficiently robust to exclude all of the spurious eigenfunctions.
There are two clear paths to address this issue.  The first is to
develop a more sophisticated filtering technique to check the
eigenfunctions during the iterative process.  The second technique
would be to use cylindrical coordinates that are better adapted to the
intense-field limit where the wavefunctions essentially are well
approximated by the product of Landau orbitals and a function of the
coordinate along the magnetic field (the so-called adiabatic
approximation).  The use of cylindrical coordinates is of course not
equivalent to the adiabatic approximation but it does fit better with
the dominant symmetry of the intense-field problem.  For this work we
did not employ cylindrical coordinates because we wanted to verify
our results against the zero-field limit where the binding energies of
the various species are known precisely.

From a mathematical point of view, we have chosen to use Chebyshev
points to sample in both the angular and radial directions.   In this
particular case it is possible to express the spectral derivatives
using a fast-Fourier transform \citep{Tref00,Boyd01}.  The most obvious
application for the problem at hand would be to speed up the
calculation of the direct component of the inter-electron potential;
because this only makes a modest contribution to the workload, we
decided to forgo the added complexity and use matrix multiplication to
determine the inter-electron potential.  On the other hand, given some
additional work it may be possible to speed up other parts of the
calculation using FFTs.

A final improvement would be to include several electronic
configurations for each state \citep[e.g.][]{Schmelcher_lithium2004}
not just the dominant one as we did here.  Given the excellent
agreement with the observed atomic properties that we found for the
zero-field limit, we expect the improvement of a multi-configuration
calculation to be modest at least for small atoms.  However, recent
theoretical results indicate that the atmospheres of neutron stars may
be composed of elements near silicon \citep{Hoff09O}, so including many
configurations may be crucial at least to verify the algorithm in the
weak-field limit.

\section{\label{sec:conclusion}Conclusion}

The amount of literature available regarding atoms in strong magnetic fields is quite extensive. Since the 1970s with improvements in both computing infrastructure and numerical methods the problem of atoms in strong magnetic fields has become increasingly tractable. However, the current state-of-the-art calculations still take on the order of hours if not days to obtain results of reasonable accuracy. To address this situation, we have, in this study, presented a new pseudospectral method for the calculation of the structure of atoms in strong magnetic fields.  The key enabling advantages of the algorithm are its simplicity (about 130 lines of
commented code) and its speed (about $10^2-10^5$ times faster than
finite-element methods to achieve similar accuracy).  For hydrogen,
helium and lithium it gives results that agree with the previous work
at the few percent level or better for fields weaker than about 
$2\times 10^{6}$~T.  

For the purpose of analyzing the spectra of magnetized white dwarfs and neutron stars it becomes crucial to have accurate data for the energy levels of different atoms in strong magnetic fields. With the ability to perform atomic structure calculations with a significant reduction in computation time, as with the pseudospectral methods described in this study, it becomes possible to amalgamate atomic structure calculation software with atmosphere models of neutron stars and white dwarf stars. This is a direct advantage of the software provided in the appendices to this paper. We have additionally presented several avenues for further research with the software including larger atoms, multi-configuration Hartree-Fock and full configuration-interaction calculations.

\section*{Acknowledgments}
This research was supported by funding from NSERC.  The calculations
were performed on computing infrastructure purchased with funds from
the Canadian Foundation for Innovation and the British Columbia
Knowledge Development Fund.  Correspondence and requests for
materials should be addressed to heyl@phas.ubc.ca.  This research has
made use of NASA's Astrophysics Data System Bibliographic Services
\bibliographystyle{mn2e}
\bibliography{math,mine,physics}

\begin{thebibliography}{}

\bibitem[\protect\citeauthoryear{Al-Hujaj \& Schmelcher}{Al-Hujaj \&
  Schmelcher}{2004}]{Schmelcher_lithium2004}
Al-Hujaj O.-A.,  Schmelcher P.,  2004, Phys. Rev. A, 70, 033411

\bibitem[\protect\citeauthoryear{Boyd}{Boyd}{2001}]{Boyd01}
Boyd J.~P.,  2001, Chebyshev and Fourier Spectral Methods.
Dover, Mineola

\bibitem[\protect\citeauthoryear{{Drake} \& {Martin}}{{Drake} \&
  {Martin}}{1998}]{1998CaJPh..76..679D}
{Drake} G.~W.~F.,  {Martin} W.~C.,  1998, Canadian Journal of Physics, 76, 679

\bibitem[\protect\citeauthoryear{Fischer}{Fischer}{1997}]{CFF1997}
Fischer C.~F.,  1997, Computational Atomic Structure, An MCHF Approach.
Institute of Physics Publishing, Bristol, UK

\bibitem[\protect\citeauthoryear{Haines \& Roberts}{Haines \&
  Roberts}{1969}]{Hain69}
Haines L.~K.,  Roberts D.~H.,  1969, Am. Journ. Phys., 37, 1145

\bibitem[\protect\citeauthoryear{Heyl \& Hernquist}{Heyl \&
  Hernquist}{1998}]{Heyl98atom}
Heyl J.~S.,  Hernquist L.,  1998, \pra, 58, 3567

\bibitem[\protect\citeauthoryear{Hoffman \& Heyl}{Hoffman \&
  Heyl}{2009}]{Hoff09O}
Hoffman K.~L.,  Heyl J.~S.,  2009, \mn, 400, 1986

\bibitem[\protect\citeauthoryear{Loudon}{Loudon}{1959}]{Loud59}
Loudon R.,  1959, Am. J. Phys., 27, 649

\bibitem[\protect\citeauthoryear{Moore}{Moore}{1971}]{Moor71}
Moore C.~E.,  1971, Technical Report~35, Atomic Energy Levels.
Natl. Bur. Stand. (U. S.) Data Ser.

\bibitem[\protect\citeauthoryear{Ralchenko, Kramida, Reader \& Team}{Ralchenko
  et~al.}{2008}]{Ralc08}
Ralchenko Y.,  Kramida A.~E.,  Reader J.,    Team N.~A.,  2008, Technical
  report, NIST Atomic Spectra Database (version 3.1.5).
National Institute of Standards and Technology, Gaithersburg, MD.

\bibitem[\protect\citeauthoryear{Ruder et~al.,}{Ruder  et~al.}{1994}]{Rude94}
Ruder H.,  et~al., 1994, Atoms in Strong Magnetic Fields : Quantum Mechanical
  Treatment and Applications in Astrophysics and Quantum Chaos.
Springer-Verlag, New York

\bibitem[\protect\citeauthoryear{Thirumalai \& Heyl}{Thirumalai \&
  Heyl}{2009}]{Thir08atom}
Thirumalai A.,  Heyl J.~S.,  2009, \pra, 79, 12514 (16 pages)

\bibitem[\protect\citeauthoryear{Trefethen}{Trefethen}{2000}]{Tref00}
Trefethen L.~N.,  2000, Spectral Methods in Matlab.
SIAM, p.~54

\end{thebibliography}
\label{lastpage}

\appendix
\begin{onecolumn}
\section{Magnetized hydrogen}
\label{sec:magnetized-hydrogen}

Both the single-electron and multiple-electron codes are self-contained {\tt
  Octave} programs with the exception of the eight-line function {\tt
  cheb} that Trefethen published previously in \citet{Tref00} and is
available online at {\tt
  http://www.comlab.ox.ac.uk/nick.trefethen/cheb.m}.  This function
calculates the differentiation matrix in Eq.~(\ref{eq:27}). Both
programs are also compatible with {\tt Matlab} with some minor
modifications in the output statements in the multi-electron program.
However, after experimentation we concluded that the {\tt Octave}
version ran faster with our setup.

The hydrogen code is written as a function to be called from the {\tt
  Octave} command line that returns the eigenfunctions, eigenvalues,
angular and radial points, a list of the eigenfunctions that pass a
rudimentary test and the ``zoom'' factor. For values of $M$ and $N$ on
the order of ten, the vast majority of the computation time is devoted to
calculating the eigenvalues, so the total execution time depends on
the efficiency of the eigenvalue routine ({\tt eig}); consequently,
porting this algorithm to C does not result in a substantial speed up
--- in fact in our experience, {\tt Octave} runs faster.
\begin{verbatim}
% HYDROGEN_S - calculate hydrogen in spherical coordinates
% beta is B/Bc, M is number of mu, N is number of radial
% the wavefunction is phi=f(r, mu=cos(theta))/r exp(i m phi)
% eigenfunctions in V, e-val in Lam, coords in (w,rs)
function [V,Lam,w,rs,igood,zoom] = hydrogen_s(beta,M,N,mphi)
% w coordinate, ranging from -1 to 1 
  [Dw,w] = cheb(M);   D2w = Dw^2;
  if (mphi~=0)  w=w(2:M);  Dw=Dw(2:M,2:M);  D2w=D2w(2:M,2:M);  end
  Hw=-diag(1-w.*w)*D2w+diag(2.*w)*Dw;
% r coordinate, ranging from 1 to -1, rp from 1 to 0
  [D,r] = cheb(N); rp = 0.5*(r+1); D = 2*D; D2 = D^2; 
  hh = diag(1-rp.*rp); D2 = hh*(hh*D2+diag(-2*r)*D);  D = hh*D;
  D2 = D2(2:N,2:N);  D =  D(2:N,2:N);  rp = rp(2:N);
% zoom factor: set by coulomb and larmor radius; , rs from inf to 0
  zoom=1/(1/110+sqrt(beta)/41);  rs=zoom*atanh(rp); R = diag(1./rs);   R2 = diag(1./(rs.*rs));   
  Hr=-1/(zoom*zoom)*D2-2*R; [rr,ww] = meshgrid(rs,w); rr = rr(:); ww = ww(:); rperp2=rr.*rr.*(1-ww.*ww);
  if (mphi==0) 
    H = kron(Hr,eye(M+1))+kron(R2,Hw)+diag(beta*beta*rperp2);
  else 
    H = kron(Hr,eye(M-1))+kron(R2,Hw)+diag(beta*beta*rperp2)+diag(mphi*mphi./rperp2); 
  end
  [V,Lam] = eig(H); Lam = diag(Lam); [Lam,ii] = sort(Lam); Lam = Lam+2*beta*(mphi-1);   V = V(:,ii);
% check outer B.C. and for bound states
  igood = find((V(1,:).*V(1,:))'<(M*N)^(-2)*1e-4 & Lam<0);
\end{verbatim}

\section{Magnetized atoms}
\label{sec:magnetized-atoms}

The atomic code is written as a script to be called from the {\tt
  Octave} command line because it has many parameters and diagnostics
that one might like to examine and the script format encourages the
user to choose what he or she wants to use.

Similar to the case of hydrogen, for values of $M$ and $N$ on the
order of ten, the vast majority of the computation time is devoted to
calculating the eigenvalues and inverting the Laplacians, so the total
execution time depends on the efficiency of the eigenvalue and matrix
inversion routines ({\tt eig} and {\tt inv}); consequently, we do not
expect that porting this algorithm to a lower-level language ({\em
  e.g.} C) will result in a substantial speed up.   In fact the
computational effort is dominated by the diagonalization that is
performed for each electron at each iteration.  The inversion of the
Laplacian is performed before the iterations begin.
\begin{verbatim}
% THREEDATOM - calculate an atom in three-D 
% the wavefunction is phi=f(r,mu=cos(theta))/r
% r coordinate, ranging from -1 to 1 
% Examples for Helium I
% nu=[1,1]; spin=[+1/2,-1/2]; mphi=[0, 0]; Z=2; %% 1s^2 1S  J=0 : 1.43189
% nu=[1,2]; spin=[-1/2,-1/2]; mphi=[0, 0]; Z=2; %% 1s2s 3S  J=1 : 1.08996
% nu=[1,2]; spin=[+1/2,-1/2]; mphi=[0, 0]; Z=2; %% 1s2s 1S  J=0 : 1.07930
% nu=[1,1]; spin=[-1/2,-1/2]; mphi=[0,-1]; Z=2; %% 1s2p 3P0 J=2 : 1.06856 
% nu=[1,3]; spin=[-1/2,-1/2]; mphi=[0, 0]; Z=2; %% 1s2p 3P0 J=1 : 1.06857 
% nu=[1,1]; spin=[+1/2,+1/2]; mphi=[0,-1]; Z=2; %% 1s2p 3P0 J=0 : 1.06856
% nu=[1,1]; spin=[-1/2,+1/2]; mphi=[0,-1]; Z=2; %% 1s2p 1P0 J=1 : 1.06593
% nu=[1,3]; spin=[-1/2,+1/2]; mphi=[0, 0]; Z=2; %% 1s2p 1P0 J=1 : 1.06594
% Examples for Lithium I
  nu=[1,1,2]; spin=[-1/2,+1/2,-1/2]; mphi=[0,0, 0]; Z=3; %% 1s^2 2s 2S  J=1/2 : 1.65595
% nu=[1,1,3]; spin=[-1/2,+1/2,-1/2]; mphi=[0,0, 0]; Z=3; %% 1s^2 2p 2P0 J=1/2 : 1.64097
% nu=[1,1,1]; spin=[-1/2,+1/2,-1/2]; mphi=[0,0,-1]; Z=3; %% 1s^2 2p 2P0 J=3/2 : 1.64097
% nu=[1,1,4]; spin=[-1/2,+1/2,-1/2]; mphi=[0,0, 0]; Z=3; %% 1s^2 3s 2S  J=1/2 : 1.62883
M=7; N=31;  MAXITER=100;  beta = 0;
% w coordinate, ranging from -1 to 1 
  [Dw,w] = cheb(M);  D2w = Dw^2;  Hw=-diag(1-w.*w)*D2w+diag(2.*w)*Dw;
% add zero weight to last term so we always use the same-sized vectors
  w_w=[inv(Dw(1:M,1:M))(1,:),0];
% r coordinate, ranging from 1 to -1, rp from 1 to 0, 
% rs from infinity to 0 (see below)
  [D,r] = cheb(N);  rp=0.5*(r+1);  D = 2*D;  D2 = D^2;
  hh=diag(1-rp.*rp);   D2=hh*(hh*D2+diag(-2*rp)*D);
  wr=inv(D(1:N,1:N))(1,:);
% drop infinite weight at infinity (i.e. 1) since w.f. goes to zero quickly
% now wr is the same dimension as D2 etc.
  wr=wr(2:N)./(1-rp(2:N).^2)';
% now multiply by hh so we didn't have zero for inversion above
  D=hh*D;
% Radial Laplacian minus r=infinity (for potential calcuations)
  D2inf = D2(2:N+1,2:N+1);
% Construct Laplacian minus r=0 and r=infinity 
% (set diricelet b.c. in radial direction for eigenvalue equations)
  D2 = D2(2:N,2:N); D =  D(2:N,2:N);  rp=rp(2:N);
  zoom=1/(1/116+sqrt(beta)/41)*0.5;  rs=zoom*atanh(rp);  wr=zoom*wr;
  R = diag(1./rs); R2 = diag(1./(rs.*rs));
% Hamilton for radial coordinate
  Hr=-1/(zoom*zoom)*D2-2*R;
  [rr,ww] = meshgrid(rs,w); rr = rr(:); ww = ww(:);  DIMEN=numel(rr);
  rperp2=rr.*rr.*(1-ww.*ww);  RR=diag(1./rr);  we=2*pi*kron(wr,w_w); 
% Full single electron Hamiltonian
  H=kron(Hr,eye(M+1))+kron(R2,Hw)+diag(beta*beta*rperp2);
  mmax=0;
  for e1cnt=1:numel(mphi)-1
    for e2cnt=e1cnt+1:numel(mphi)
      if (spin(e1cnt)==spin(e2cnt)) 
        dum=abs(mphi(e1cnt)-mphi(e2cnt));
        if (dum>mmax) mmax=dum; end
      end
    end
  end
% Construct Laplacians for electron potential calculation
  Linv=zeros(DIMEN,DIMEN,mmax+1);
  for i=0:mmax
% set the denominator equal to one, if it goes to zero (don't want an infinity)
% we will drop rperp2=0 later
    L=kron(1/(zoom*zoom)*D2,eye(M+1))-kron(R2,Hw)-diag(i*i./(rperp2+(rperp2==0)));
% set the BC for rperp2=0 but add in identity so can be inverted
% an infinity in the last step would have given an NaN here 
    if (i~=0) L=diag(rperp2==0)+diag(rperp2~=0)*L*diag(rperp2~=0); end
% add in the origin as a single point
    L=[ [L, kron(D2inf(1:N-1,N),ones(M+1,1))]; kron(D2inf(N,1:N-1),ones(1,M+1)), D2inf(N,N)];
% invert Laplacian and set boundary condition at origin
    Linvhold=(eye(DIMEN+1)-[zeros(DIMEN+1,DIMEN),ones(DIMEN+1,1)])*inv(L);
% divide by R before and after and remove origin!
% don't need to set B.C. again because the solutions for mphi~=0 are already zero in the 
% right places, and the e-val equation sets them to zero when needed
    Linv(:,:,i+1)=-4*pi*RR*Linvhold(1:DIMEN,1:DIMEN)*RR;
  end
  u=zeros(DIMEN,numel(nu));
  direct_old=zeros(DIMEN,DIMEN,numel(nu));
  exchange_old=zeros(DIMEN,DIMEN,numel(nu),mmax+1);
  direct_new=zeros(DIMEN,DIMEN,numel(nu));
  exchange_new=zeros(DIMEN,DIMEN,numel(nu),mmax+1);
  interact=zeros(DIMEN,DIMEN);
% begin iterations
  etotold=1e33;  etot=1;  j=1;
  while (abs((etot-etotold)/etot)>1e-5 & j<MAXITER) 
% calculate each electron wf
    etotold=etot;
    etot=0;
    for e1cnt=1:numel(nu)
      interact=zeros(DIMEN,DIMEN);
      for e2cnt=1:numel(nu)
        if (e2cnt~=e1cnt) 
          if (spin(e1cnt)==spin(e2cnt)) 
            interact=interact+direct_old(:,:,e2cnt) ...
                -exchange_old(:,:,e2cnt,abs(mphi(e1cnt)-mphi(e2cnt))+1);
          else
            interact=interact+direct_old(:,:,e2cnt);
          end
        end
      end
      if (mphi(e1cnt)==0) 
        [V,Lam] = eig(H+2/Z*interact);
      else
        Htmp=H+2/Z*interact+diag(mphi(e1cnt)*mphi(e1cnt)./rperp2);
% apply boundary condition along z-axis, remove columns and rows with rperp2==0
        [V,Lam] = eig(reshape(Htmp(find(rperp2*rperp2'~=0)),(N-1)*(M-1),(N-1)*(M-1)));
      end
      Lam = diag(Lam);  [Lam,ii] = sort(real(Lam)); 
      igood = find((V(1,:).*V(1,:))'<(M*N)^(-2)*1e-4);
      V = V(:,ii);   uu=V(:,igood(nu(e1cnt)));
% add in boundary condtion for angular direction (along z-axis), if needed
      if (mphi(e1cnt)~=0) 
        uu=[zeros(1,N-1);reshape(uu,M-1,N-1);zeros(1,N-1)](:); 
      end
      l(e1cnt)=Lam(igood(nu(e1cnt)))+2*beta*(mphi(e1cnt)+spin(e1cnt)-0.5);
      u2=uu.*uu; norm=we*u2; u2=u2/norm; uu=uu/sqrt(norm); u(:,e1cnt)=uu; % normalize
      eeen(e1cnt)=2/Z*we*(uu.*(interact*uu));  etot=etot+l(e1cnt)-eeen(e1cnt)/2;
      printf('J=%d e1=%d E-val: %18.12f EE: %18.12f Etot: %18.12f\n',j,e1cnt,l(e1cnt),eeen(e1cnt),etot);
      direct_new(:,:,e1cnt)=diag(Linv(:,:,1)*u2);
      for i=1:mmax+1 exchange_new(:,:,e1cnt,i)=diag(uu)*Linv(:,:,i)*diag(uu); end
    end
    j=j+1;  direct_old=direct_new;  exchange_old=exchange_new;
 end
 if (j==MAXITER) 
   printf('Did not converge\n');
 end

\end{verbatim}
\end{onecolumn}

\end{document}